\documentclass[aps,pra,notitlepage]{revtex4}
\usepackage[dvips]{graphicx}
\usepackage{amsmath,amssymb,enumerate,color,ulem,url}
\usepackage{subfigure}
\newcommand{\bra}[1]{{\left\langle #1 \right\vert}}

\newcommand{\ket}[1]{{\left\vert #1 \right\rangle}}

\begin{document}
\title{Decoherence of Nitrogen Vacancy Centers in Diamond}
\author{Shigeru Ajisaka}
\email{g00k0056@suou.waseda.jp}
\author{Y. B. Band}
\email{band@bgu.ac.il}
\affiliation{
Department of Chemistry, Department of Physics, Department of Electro-Optics, and the Ilse
Katz Center for Nano-Science, \\
Ben-Gurion University, Beer-Sheva 84105, Israel, and\\
New York University and the NYU-ECNU Institute of Physics at NYU Shanghai, 3663 Zhongshan Road North, Shanghai, 200062, China}

\begin{abstract}
We model the decoherence and dephasing of nitrogen vacancy (NV) centers in diamond due to a noisy paramagnetic bath, with and without the presence of a rf field that couples levels of the ground electronic state manifold, using a simple quantum mechanical model that allows for analytical solutions.  The model treats the NV three-level ground state system in the presence of fluctuating magnetic fields that arise from the environment, and that result in decoherence, dephasing and dissipation.  We show that all 9 eigenmodes of the three-level system are coupled to each other due to interaction with the environment, and we discuss consequences for fitting experiments in which decoherence plays a role.
\end{abstract}

\maketitle

\section{Introduction}

Nitrogen-vacancy (NV) color centers in diamond have become platforms for quantum information processing because the spin of the NV can be manipulated~\cite{Gruber97} and is easy to initialize to the ground state level, even at room temperatures~\cite{Dutt07, Jacques09}.  NVs have a long relaxation time at room (and lower) temperatures, hence single spin control can be achieved over long time scales~\cite{Mizuochi09}.  Moreover, the quantum state of NVs can be optically read out~\cite{Oort88, Jiang08, Neumann10}.  Furthermore, Gruber {\it et al.}~\cite{Gruber97} have detected and manipulated an individual NV center.  In quantum information and computation applications, controllability of the quantum state in a solid-state medium is very desirable.  Entanglement of two NV centers separated by 10 nm was reported in Ref.~\cite{Dolde13} and entanglement of NVs in two separate diamond samples with a spatial separation of three meters was obtained in Ref.~\cite{Bernien13}.  NVs are also good candidates for many technological applications, such as quantum gates~\cite{Jelezko04}, quantum cryptography~\cite{Krutsiefer00, Beveratos02}, magnetic sensing~\cite{Maze08,Balasubramanian09}, electric field sensing~\cite{Dolde11}, temperature sensing~\cite{Neumann13,Kucsko13} and sensing of inertial effects~\cite{Ledbetter12}.  Maze reported a magnetic sensor and measured 0.5 $\mu$T magnetic fields with about ten nanometer spatial resolution~\cite{Maze08}. Because NVs can be used at room temperature, they can be used for magnetic sensing in biological systems. Indeed, the Walsworth group visualized a live bacteria with 400 nm spatial resolution~\cite{Sage13}.  However, the effects of the environment on the system often remains an obstacle to achieving better performance. Environmental effects have been experimentally studied and have been discussed theoretically. For example, the importance of hyperfine interactions have been studied using electron spin echo modulation spectroscopy~\cite{Sloop81, Oort90}, Raman-heterodyne spectroscopy~\cite{Xing93} and Ramsey fringe spectroscopy~\cite{Childress06}, and has also been studied theoretically~\cite{Maze08PRB}.  The temperature dependence of relaxation times has been experimentally studied~\cite{Reynhardt98, Jarmola12, Mrozek15}, and the effects of phonons has been treated~\cite{Bennett13}.  Understanding the effects of the environment, i.e., the effects of other spins, phonons, and other nearby impurities in the crystal lattice, on the electronic and spin degrees of freedom of the NV center is imperative in order to control environmental effects and minimize the adverse effects of noise and decoherence.

The decay dynamics of spin systems is typically characterized by three decay times: (1) The relaxation time of the population towards equilibrium; this time scale is denoted as $T_1$. (2) The relaxation time of the coherence of a single spin; this time scale is denoted as $T_2$. (3) The relaxation time for the coherence of an ensemble of spins~\cite{Clarke08}.  If each spin is placed in a  slightly different environment, the resonance frequency has a width whose inverse corresponds to a decay time called $T_2^*$.
Hence, even if each spin maintains coherence, the sum of many signals shows what is known as free induction decay.

For a spin 1/2 system, the dynamics of the spin can be mapped to a three dimensional Bloch sphere and the two relevant decay times of a measurement of a single spin are $T_1$ and $T_2=T_2^*$.  Increasing $T_1$ and $T_2$ is important for improving quantum coherence, and therefore vital for quantum information and computing applications.  Experimentally, $T_2$ can be measured for an ensemble of spins by removing nonuniform effects using spin echo techniques; a longitudinal spin relaxation time of $T_2 \approx 0.6$ s was obtained for NVs at 77 K using dynamical decoupling techniques~\cite{Bar-Gill13}.  But even longer coherence time is hoped for.  For a spin 1/2 system, a three dimensional Bloch sphere picture precisely captures the spin physics, and the only decay rates required are $T_1, T_2$ and $T_2^*$.  However, an {\it ab initio} study of the electronic structure of NV centers~\cite{Gali08, Lenef96} reveals that the  ground state is a spin {\it triplet} with a zero magnetic field splitting ${\cal D} = 2.87$ between the $m=0$ level and the $m = \pm 1$ levels.  This is confirmed by experiment~\cite{Loubser77, Reddy87}.  Nevertheless, when transitions  between $m=0$ and $m= \pm 1$ are selected, typically the dynamics is analyzed in terms of a three dimensional Bloch sphere. However, there is always inevitable leakage to the other level through the interaction with the environment.  Thus, in general, 9 eigenmodes (one of which, the steady state mode, always has zero eigenvalue) are required to fully account for the decoherence of a spin triplet system (see below).

Here we study a single negatively charged diamond NV center, NV$^{-}$, using a simple quantum mechanical model, which allows for analytical solutions, hence, our results yield simple physical insights for understanding relaxation processes in generic three level systems.  More specifically, we consider a three-level system in the presence of fluctuating magnetic fields arising due to its environment and that results in decoherence, dephasing and dissipation.  We calculate the dynamics with and without the presence of a radio frequency field that can strongly couple the triplet levels (e.g., $m=1$ and $m=0$).  We show that all 9 eigenmodes of the system are coupled to each other due to interaction of the NV with its environment.  Hence, dynamics observed in NV experiments generally result due to combinations of all eigenmodes.  We further fit the evolution of a triplet spin system with a smaller number of decay times (that are generally somewhat different from the genuine 9 eigenmodes) and the fits are of excellent quality. This has practical ramifications regarding fitting experimental data.  We discuss the dynamics the presence of a radio frequency field which strongly couples the $m=1$ and $m=0$ levels, and without such a field.

This paper is organized as follows.  Section~\ref{Sec:Model} introduces a model of an NV center in the presence of a fluctuating magnetic field. In Sec.~\ref{Sec:wo_Strain} we analyze a case without strain or electric field effects ($\epsilon=0$, see below). Section~\ref{Sec:w_Strain} considers the case with strain.  In Sec.~\ref{Sec:rf} we discuss the dynamics in the presence of a radio frequency field that can couple the levels of the NV, and Sec.~\ref{Sec:Decoherence} considers coherence relaxation (transverse spin relaxation), i.e., the decoherence of off-diagonal elements of the density matrix.  Section~\ref{Sec:summary} contains a summary and conclusion.

\section{The Model}  \label{Sec:Model}

The symmetry group of the NV center is $C_{3v}$ and the ground state of NV$^{-}$ is a ${}^3 A_2$ state, where this state designation refers to an irreducible representation of the $C_{3v}$ group.  The ground state manifold of the NV in the presence of an external magnetic field can be modeled in terms of a spin 1 Hamiltonian.  The deterministic part of the Hamiltonian (the part without a stochastic magnetic field) is given by \cite{Doherty13}
\begin{eqnarray}  \label{Eq:H_nv}
	H&=& {\cal D} J_z^2 + \mu \, {\bf J} \cdot {\bf B} +
	\epsilon_x (J_x^2-J_y^2)+ \epsilon_y (J_x J_y-J_y J_x)    \nonumber \\
	&=&
		\left(
		\begin{array}{ccc}
			{\cal D}+\mu B_z & 0 & \epsilon^*\\
			0 & 0 & 0\\
			\epsilon & 0 & {\cal D}-\mu B_z	,				
		\end{array}
		\right) ,
\end{eqnarray}
where the diagonal matrix elements of $H$, $\omega_1={\cal D}+\mu B_z$, $\omega_2=0$, and $\omega_{-1}={\cal D}-\mu B_z$, incorporate ligand field effects and the Zeeman interaction with the deterministic magnetic field, with the zero magnetic field splitting parameter, ${\cal D} = 2.87$ GHz, and the magnetic moment of the NV, $\mu = 1.40$ MHz/G.  For simplicity, we have taken the deterministic magnetic field to be oriented along the $z$-axis (the axis of the NV center).  The angular momentum components $J_i$ ($i = x, y$ and $z$) are spin 1 matrices \cite{Band_Avishai}, and $\epsilon \equiv \epsilon_x + i\epsilon_y$ is the complex strain coefficient, which also incorporates the effects of an external electric field (if present) \cite{Doherty13}.  

An additional time-dependent Zeeman spin 1 Hamiltonian, $H_{\rm{st}}(t) =  \mu \, {\bf J} \cdot {\bf b}_{\rm{st}}(t)$, that contains a fluctuating (stochastic) magnetic field ${\bf b}_{\rm{st}}(t)$ due to the presence of other magnetic moments that comprise the local environment experienced by the NV can be added to the Hamiltonian in Eq.~(\ref{Eq:H_nv}).  Thus, the full Hamiltonian becomes
\begin{eqnarray}  \label{Eq:H_nv2}
	H(t) &=& H + H_{\rm{st}}(t) =
		\left(
		\begin{array}{ccc}
			{\cal D}+\mu B_z & 0 & \epsilon^*\\
			0 & 0 & 0 \\ 
			\epsilon & 0 & {\cal D}-\mu B_z	,				
		\end{array}
		\right) +\sum_i  \sqrt{\gamma_i} \xi_i(t) J_i ,
\end{eqnarray}
where $\sqrt{\gamma_i}$ is the strength of the magnetic field fluctuation and $\xi_i(t)$ is proportional to the normalized [see (\ref{B_fluc})] magnetic field fluctuation along the $i$th axis at the NV location ($i=x,y,z$).

For the case of Gaussian white noise in the magnetic field components, i.e., 
\begin{eqnarray} \label{B_fluc}
\langle \xi_i(t)\rangle=0,\ \  \langle \xi_i(t) \xi_j^*(t') \rangle = \delta_{i,j} \delta(t-t'),
\end{eqnarray}
the Schr\"{o}dinger equation with the presence of noise should have the following form in order to guarantee the unit normalization of the realization average of the norm $\| \ket{\psi (t)} \|$ \cite{vanKampenBook},
\begin{equation}  \label{Stoch_Sch_Eq}
\frac{d}{dt} \ket{\psi (t)}= -i H(t)\ket{\psi (t)} - \sum_i \frac{\gamma_i}{2} J_i^2\ket{\psi(t)} .
\end{equation}
For Gaussian white noise, it is well known that the realization average can be obtained using a master equation with Lindblad operators ${\cal L}_i$ multiplied by Lindblad coefficients $\gamma_i$ where $(i=x,y,z)$ are constants.  The generic form of the Liouville-von Neumann equation for the density matrix with Lindblad operators ${\cal L}_i$ is \cite{vanKampenBook, BreuerBook, SchlosshauerBook},
\begin{equation}
\frac{d\rho}{dt}=-i[H,\rho]+\sum_{i,j} L_{i,j} \left( {\cal L}_i \rho  {\cal L}_j^\dag-\frac{1}{2}\{ {\cal L}_j^\dag  {\cal L}_i,\rho\}\right) ,
\end{equation}
where $\{L_{i,j}\}$ is a positive coefficient matrix \cite{Lindblad76, Gorini76}.
In our case, where the stochastic magnetic field components are coupled to the angular momentum operators through the stochastic Zeeman Hamiltonian, $H_{\rm{st}}(t) = \mu \, {\bf J} \cdot {\bf b}_{\rm{st}}(t)$, the stochastic magnetic field components yield Lindblad operators
\begin{equation}
 {\cal L}_i = J_i ,
\end{equation}
and Lindblad coefficients can be taken to be $\gamma_i$ where $(i=x,y,z)$, i.e., the Lindblad matrix $\{L_{i,j}\}$ is a diagonal matrix with nonzero matrix elements, $L_{xx}=\gamma$, $L_{yy}=\gamma$, and $L_{zz}=\Gamma \ne \gamma$ (at least not necessarily equal to $\gamma$).

The eigenvalues of the Liouvillian operator in the Liouville-von Neumann equation for the density matrix of the system determine the dynamics.  Each of the matrix elements of the density matrix can be expressed as
\begin{equation}  \label{Eq:rho_expand}
\rho_{\alpha\beta}(t) =a^{\alpha,\beta}_0 +\sum_{i=1}^8 a_i^{\alpha,\beta} \exp (\lambda_i t) ,\ \ 
(\alpha,\beta=1,2,3)
\end{equation}
where the $\lambda_i$ ($i = 0, 1, \ldots, 8$) are the 9 eigenvalues of the 9$\times$9 Liouvillian operator [henceforth, we call the $\lambda_i$ the generalized eigenmode frequencies \cite{NoteRapidity}, or eigenmodes for short], $a_i$ ($i = 1, \ldots, 8$) are the amplitude coefficients, and the coefficient $a_0$ corresponds to the amplitude of the steady state whose existence is guaranteed by the trace preserving property.  The eigenmode $\lambda_0 = 0$ (hence $\exp (\lambda_0 t) = 1$).  As seen in the following sections, some of the coefficients $a_i$ may be zero if the Hilbert space is separated due to a special symmetry of the Liouvillian.

Finally, we remark that, as opposed to our case where the angular momentum operators $J_i$ are chosen to be the Lindblad operators, Lindblad operators of the form $c^\dag_\alpha c_\beta$ do {\it not} mix off-diagonal elements of the density matrix.  Such Lindblad operators cause dephasing and transitions between {\it diagonal} elements~\cite{Bar-Gill13}. Consequently, these two cases are quite different.  When $c^\dag_\alpha c_\beta$ is used as a Lindblad operator, it causes the population dynamics of diagonal elements from $\rho_{\beta\beta}$ to $\rho_{\alpha\alpha}$ as well as the dephasing of four density matrix elements ($\rho_{\alpha \beta}$, $\rho_{\beta \overline{\alpha \beta}}$ and their conjugates, where $\overline{\alpha\beta}$ is an index that is not equal to $\alpha$ nor $\beta$) without mixing the density matirx elements. 

Choosing angular momentum operators, e.g., $J_x= (c_1^\dag c_0 + c_0^\dag c_1 + c_0^\dag c_{-1}+ c_{-1}^\dag c_{0})/2$, as Lindblad operators results in very different behavior than choosing the four transition operators $c_1^\dag c_0$, $c_0^\dag c_1$, $c_0^\dag c_{-1}$, and $c_{-1}^\dag c_0$ as Lindblad operators.  For example, the first term of the Lindblad dissipator associated with an angular momentum $J_x$, i.e., $J_x \rho J_x$, can be written as
\begin{eqnarray*}
\frac{1}{2} (c^\dag_1 c_0 + c^\dag_0 c_1 + c^\dag_0 c_{-1} + c^\dag_{-1} c_0) \rho(c^\dag_1 c_0 + c^\dag_0 c_1 + c^\dag_0 c_{-1} + c^\dag_{-1} c_0), 
\end{eqnarray*}
whereas the first term of the Lindblad dissipator associated with the four transition operators contains only four terms:
\begin{eqnarray*}
c^\dag_1 c_0 \rho c^\dag_0 c_1
+c^\dag_0 c_1 \rho c^\dag_1 c_0
+c^\dag_0 c_{-1} \rho c^\dag_{-1} c_0
+c^\dag_{-1} c_0 \rho c^\dag_0 c_{-1} .
\end{eqnarray*}
Terms such as $c^\dag_1 c_0\rho c^\dag_0 c_{-1}$ appear only if angular momentum is chosen as a Lindblad operator.  This kind of term mixes the diagonal elements and the off-diagonal elements of the density matrix.  On the other hand, terms which appear in both cases only affect populations (the diagonal elements of the density matrix) , i.e., the term $c^\dag_i c_j\rho (c^\dag_i c_j)^\dag$ increases the population of state $i$, $\rho_{ii}$.  Linblad operators involving transition operators yield population rate equations ($\frac{d}{dt} \rho_{ii} = \ldots)$, whereas Linblad operators involving angular momentum operators yield master equations with two sets of closed equations; one set for the diagonal elements that are coupled to $\rho_{1-1}$ and $\rho_{-11}$, and one set for the other off-diagonal matrix elements, $\rho_{10}, \rho_{01},\ \rho_{0-1}$ and $\rho_{-10}$.  Note that $\{(c^\dag_i c_j)^\dag c^\dag_i c_j,\rho \}$ terms yield the decay of off-diagonal elements density matrix elements $\rho_{ij}$ as well as the diagonal density matrix element $\rho_{jj}$.

To summarize, the angular momentum Lindblad operators model decoherence and dissipation due to random magnetic fields arising from the presence of magnetic impurities that make up a noisy paramagnetic bath.  The Lindblad operators corresponding to transition operators can be used to model the asymmetric transition between two levels and are typically used to obtain local detailed balance of two-levels system.

Before starting our detailed analysis, let us comment on the relation between the bound on coherence times and the Lindblad equation for two-level systems.  From experiments that measure $T_1$ and $T_2$, it is well known that the coherence time $T_2$ is bound by twice the population decay time, $2 \, T_1$, for two-level systems. This relation is closely related to the Lindblad equation; the relation $1/\tau_1 + 1/\tau_2 + 1/\tau_3 \ge 2 /\tau_i$, where $1/\tau_i$ is the real part of the $i$th eigenmode, was proved for any Lindbladian time evolution~\cite{Gorini76, Kimura02}. Eigenmodes of two-level systems are either three real values or have one real and a pair of complex conjugate values.  The later is observed in most systems because the Hamiltonian yields the imaginary parts.  Taking $\tau_2 = \tau_3$ casts the inequality into the form $2 \tau_1 \ge \tau_2$.  However, to the best of our knowledge, the extension of this relation to $N$-level systems is still an open question, but we expect that a similar inequality may hold for $N$-level systems.  We note that a stronger limit was reported for NVs, i.e., $T_2$ is limited to approximately $T_1/2$ over a wide range of temperatures~\cite{Bar-Gill13}.

\section{Equations of Motion Without Strain ($\epsilon=0$)}   \label{Sec:wo_Strain}

In this section, we study a case without strain and without the presence of an external electric field, i.e., $\epsilon=0$.  In this case, we observe
\begin{itemize}

\item The density matrix elements $\rho_{10}$ and $\rho_{0-1}$ do not couple to their conjugates, i.e., $\rho_{01}$ and $\rho_{-10}$. 	Thus, one only needs to solve 2$\times$2 matrix problem rather than  4$\times$4, which is the special property for non-strain case $\epsilon=0$.

\item Diagonal elements are decoupled from the off-diagonal elements~$\rho_{1-1}$ and $\rho_{-11}$.
Therefore one can model the relaxation rate of the diagonal elements and that of the off-diagonal elements independently since they decoupled in this case.  However, as we shall show below, this is not the case if strain or a rf field is present. 
\end{itemize}

We first discuss the diagonal elements of the density matrix.  The dynamics is determined by the differential equations
\begin{eqnarray*}
\frac{d}{dt} \rho_{00} &=& \gamma(1-3{\rho}_{00}) ,
\\
\frac{d}{dt} \Delta\rho &=& -\gamma \Delta\rho ,
\end{eqnarray*}
where $\Delta\rho(t) = {\rho}_{-1 \, -1} (t) - {\rho}_{11} (t)$.
The population $\rho_{00}(t)$ and $\Delta\rho(t)$ are given explicitly by 
\begin{eqnarray*}
	\rho_{00}(t) &=& \frac{1}{3}+\left({\rho}_{00}(0)-\frac{1}{3}\right) \exp(-3\gamma t) ,\\
	\Delta \rho(t) &=&  \Delta \rho(0) \exp(-\gamma t) .
\end{eqnarray*}
Equivalently, $\rho_{\pm 1\pm 1}(t)$ is given by
\begin{eqnarray*}
 \dot{\rho}_{\pm 1\pm 1}(t) &=& \frac{1}{3}-\frac{1}{2}\left({\rho}_{00}(0)-\frac{1}{3}\right) \exp(-3\gamma t)
 \pm\frac{\rho_{11}(0)-\rho_{-1 \, -1}(0)}{2} \exp(-\gamma t) .
\end{eqnarray*}

Next, we present the solution for the off-diagonal elements. $\rho_{1-1}(t)$ is determined by solving
\begin{eqnarray*}
 \dot{\rho}_{1-1}(t) = -Z \rho_{1-1} ,
\end{eqnarray*}
where $Z=(\gamma+2\Gamma) + i \omega_{1-1}$ and $\omega_{ij}\equiv \omega_i-\omega_j$, hence, the solution is 
\begin{eqnarray*}
\rho_{1-1}=C \exp\left[ (-(\gamma+2\Gamma)-i \omega_{1-1}) t\right] .
\end{eqnarray*}
The density matrix elements $\rho_{10}$ and $\rho_{0-1}$ can be obtained from
\begin{eqnarray}  \label{Eq:r_off}
\frac{d}{dt}
		\left(
		\begin{array}{c}
			\rho_{10} \\
			\rho_{0-1}
		\end{array}
		\right)
		=
		\left(
		\begin{array}{cc}
			-\tilde{\Gamma} -i \omega_{10} & \gamma \\
			\gamma & -\tilde{\Gamma}+i\omega_{-10}  
		\end{array}
		\right)
		\left(
		\begin{array}{c}
			\rho_{10} \\
			\rho_{0-1}
		\end{array}
		\right) .
\end{eqnarray}
Defining the discriminant of the characteristic polynomial as $D \equiv 4\gamma^2-(\omega_1-2\omega_0 +\omega_{-1})^2=4[\gamma^2-(\mu B_z)^2]$, the eigenvalues of the matrix on the right hand side of Eq.~(\ref{Eq:r_off}) are easily obtained.

The relaxation rate for the non-strain case, i.e.,  $\epsilon=0$, is summarized as follows (provided ${\cal D}>\gamma$):
\begin{itemize}
	\item[(1)]  Steady state is an equally distributed diagonal state, i.e., $\rho_{{\mathrm SS}}=\frac{1}{3}\left( \ket{1}\bra{1} +\ket{-1}\bra{-1}+ \ket{0}\bra{0} \right)$.
	\item[(2)]  Relaxation rate for $\ket{0}\bra{0}$ is $3\gamma$: $\rho_{00}(t) = \frac{1}{3} +\left({\rho}_{00}(0)-\frac{1}{3}\right) \exp(-3\gamma t)$. 
	\item[(3)]  Relaxation rate between $\ket{1}\bra{1}$ and $\ket{-1}\bra{-1}$ is $\gamma$:
$\Delta \rho(t) =  \Delta \rho(0) \exp(-\gamma t)$. 
	\item[(4)]  Relaxation rate for $\ket{1}\bra{-1}$ is $\gamma+2\Gamma$. This element shows an oscillation with frequency $2 \mu B_z$.
	\item[(5)]  Relaxation rate for $\ket{\pm 1}\bra{0}$  is 
	$\tilde{\Gamma}\equiv (3\gamma+\Gamma)/2$.
	This element contains oscillations with frequencies $\mu B_z \pm \sqrt{{\cal D}^2-\gamma^2}$. 	Since the transition between $m=\pm 1$ is almost forbidden, $T_2$ measurement measures this decay rate.
\end{itemize}

\begin{figure}
\begin{center}
\includegraphics[width=0.55\textwidth]{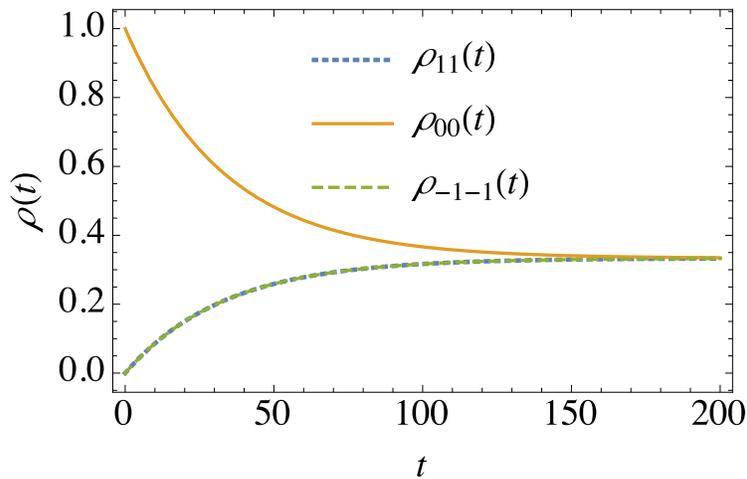}
\caption{(Color online) Population decay for the case with $\epsilon=0$: $\rho_{11}(t)$ (blue dotted curve), $\rho_{00}(t)$ (orange), $\rho_{-1 -1}(t)$ (green dashed curve).}
\label{Fig:Decay}
\end{center}
\end{figure}

Figure~\ref{Fig:Decay} shows the decay of the diagonal matrix elements of $\rho(t)$ as a function of time for $\mu B_z = 1$ GHz and $\rho_{00}(0) = 1$.

\section{Equations of Motion With Strain} \label{Sec:w_Strain}

In this section we study the case with strain, $\epsilon\neq 0$.  The equation of motion for ${\rho}_{00}$(t) remains the same as in the non-strain case~$\epsilon=0$.  The steady state is ${\rho}_{00} = 1/3$, hence, in steady state ${\rho}_{11}+{\rho}_{-1 \, -1}=2/3$.  If there is symmetry between ${\rho}_{11}$ and ${\rho}_{-1 \, -1}$, one finds $\Delta \rho = 0$, i.e., ${\rho}_{11}={\rho}_{-1 \, -1}=1/3$.

The equation of motion for $\Delta\rho$, $\rho_{1-1}$, and $\rho_{-11}$ is
\begin{eqnarray}
\frac{d}{dt}
		\left(
		\begin{array}{c}
			\Delta\rho\\
			\rho_{1-1} \\
			\rho_{-11} 
		\end{array}
		\right)
		=
		\left(
		\begin{array}{ccc}
			-\gamma & -2i\epsilon^* & 2i \epsilon \\
			-i\epsilon & -Z & 0 \\
			i\epsilon^* & 0 & -Z^* 
		\end{array}
		\right)
		\left(
		\begin{array}{c}
			\Delta\rho\\
			\rho_{1-1} \\
			\rho_{-11} 
		\end{array}
		\right)
\equiv
A
		\left(
		\begin{array}{c}
			\Delta\rho\\
			\rho_{1-1} \\
			\rho_{-11} 
		\end{array}
		\right)	.
\end{eqnarray}
The matrix $A$ is invertible because the determinant of the matrix $A$, ${\mathrm{det}}[A] = -\gamma (\gamma+2\Gamma)^2 - 4(\gamma+ 2\Gamma) |\epsilon|^2- \gamma \omega_{1-1}^2 < 0$, since $\gamma$ and $\Gamma$ must be non-negative. Therefore, quite generally, at steady state, $\Delta\rho=\rho_{1-1}-\rho_{-11} = 0$, i.e., $\rho_{1-1} = \rho_{-11} = \rho_{00} = 1/3$.

The eigenvalues of the matrix $A$ for $\epsilon_x = \epsilon_y \equiv \epsilon_s$, where the subscript $s$ stands for symmetric, are given by
\begin{eqnarray*}
\lambda_1 &=& \frac{1}{3}
	\left(
		F-(3\gamma+4\Gamma)+\frac{4\Gamma^2-3(8\epsilon_s^2+\omega_{1-1}^2)}{F}
	\right) ,
\\
\lambda_2 &=& \frac{1}{6}
	\left(
		(-1+\sqrt{3} i)F-2(3\gamma+4\Gamma)-(1+\sqrt{3}i)\frac{4\Gamma^2-3(8\epsilon_s^2+\omega_{1-1}^2)}{F}
	\right) ,
\\
\lambda_3 &=& \lambda_2^* ,
\end{eqnarray*}
where
$$F^3 = 8\Gamma^3 + 18\Gamma(\omega_{1-1}^2-4\epsilon_s^2) + 3\sqrt{3}\sqrt{ 16 \omega_{1-1}^2 \Gamma^4-8(8\epsilon_s^4+20\epsilon_s^2\omega_{1-1}^2-\omega_{1-1}^4)\Gamma^2 + (8\epsilon_s^2+\omega_{1-1}^2)^3} .$$
Hence, the relaxation rates are $-\lambda_1$ and -Re$[\lambda_2]$.
Note that in the case without strain ($\epsilon=0$), $\Delta\rho$ and the off-diagonal elements, $\rho_{1-1}$, $\rho_{-11}$ are decoupled. Therefore, the first eigenvector is expected to be localized to the diagonal elements, while the latter two eigenvectors are expected to be localized to the off-diagonal elements.  
For example, taking $\gamma = \Gamma = 0.01$, $\epsilon_s = 0.01$, and $\omega_{1-1} = 0.4$, 
gives $\lambda_1 = -0.0100993$, $\lambda_2 = -0.0299504 + 0.400996 \, i$, and $\lambda_3 = -0.0299504 + 0.400996 \, i$.  
The eigenvectors $\vec{v}_1$,  $\vec{v}_2$, and $\vec{v}_3$ are localized in the first, second, and third elements respectively.  Therefore, $-\lambda_1$ is an approximate relaxation rate for $\Delta\rho$, and -Re[$\lambda_2$] is an approximate relaxation rate for $\rho_{1-1}$ and $\rho_{-11}$.

The other elements of the density matrix satisfy the differential equation
\begin{eqnarray}  \label{Eq:rho_4_4}
		\left(
		\begin{array}{c}
			\rho_{1 0} \\
			\rho_{0 \, -1} \\
			\rho_{0 1} \\
			\rho_{-1 0} 
		\end{array}
		\right)
		=
		\left(
		\begin{array}{cccc}
			-\tilde{\Gamma} -i \omega_{10} & \gamma & 0 & -i\epsilon\\
			\gamma & -\tilde{\Gamma} +i \omega_{-10}  & i\epsilon &0\\
			0 & i\epsilon^* &-\tilde{\Gamma}  +i \omega_{10} & \gamma\\
			-i\epsilon^* & 0 & \gamma &-\tilde{\Gamma}  -i \omega_{-10} 
		\end{array}
		\right)
		\left(
		\begin{array}{c}
			\rho_{1 0} \\
			\rho_{0 \, -1} \\
			\rho_{0 1} \\
			\rho_{-1 0} 
		\end{array}
		\right) ,
\end{eqnarray}
where $\tilde{\Gamma} \equiv \frac{3}{2}\gamma+\frac{\Gamma}{2}$.  The matrix on the right hand side of Eq.~(\ref{Eq:rho_4_4}) has the following four eigenvalues:
\begin{eqnarray}
\lambda &=& -\tilde{\Gamma}\pm \frac{i}{\sqrt{2}} \sqrt{ -(2\gamma^2 - 2|\epsilon|^2-\omega_{10}^2-\omega_{-10}^2) \pm
		\sqrt{ -4\gamma^2(\omega_{10}-\omega_{-10})^2+(4|\epsilon|^2+(\omega_{10}-\omega_{-10})^2) (\omega_{10}+\omega_{-10})^2 }} , 
		\nonumber \\ 
&=&  -\tilde{\Gamma}\pm	i\sqrt{(\mu	B_z)^2+{\cal D}^2-\gamma^2 + |\epsilon|^2 \pm 2\sqrt{(\mu B_z)^2({\cal D}^2-\gamma^2)+{\cal D}^2 |\epsilon|^2}} .
\label{eigenvalue4}
\end{eqnarray}
The necessary conditions for the eigenvalues to be of the form $-\tilde{\Gamma}$ + (a pure imaginary number) are:
\begin{eqnarray*}
&&(\mu B_z)^2+({\cal D}^2-\gamma^2) + |\epsilon|^2>0 , \\
&&(\mu B_z)^2({\cal D}^2-\gamma^2)+{\cal D}^2 |\epsilon|^2 >0 , \\
&&\left(\mu B_z)^2-\sqrt{{\cal D}^2-\gamma^2}\right)^2+|\epsilon|^2 
	\left\{2({\cal D}^2-\gamma^2)+2(\mu B_z)^2 - 4{\cal D}^2+ |\epsilon|^2 \right\}>0 .
\end{eqnarray*}
Thus, if $\epsilon \ll {\cal D}$ and $|\mu B_z|$, and $\gamma < {\cal D}$, the relaxation rate is same as the non-strain case, $\tilde{\Gamma} = \frac{3}{2}\gamma +\frac{\Gamma}{2}$.

We note that, with the aid of the equality
\begin{eqnarray*}
\sqrt{(\mu B_z)^2+{\cal D}^2-\gamma^2 + |\epsilon|^2 \pm 2\sqrt{(\mu B_z)^2({\cal D}^2-\gamma^2)+{\cal D}^2 |\epsilon|^2}}
\Bigg|_{\epsilon=0}= 
\left|\mu B_z\pm \sqrt{{\cal D}^2-\gamma^2}\right| ,
\end{eqnarray*}
the eigenvalues expressed in Eq.~(\ref{eigenvalue4}) are consistent with the non-strain case.

In summary, for non-vanishing strain, $\epsilon\neq 0$, the steady state is the same as for the no-strain case.  Even though $\rho_{10}$ and $\rho_{0-1}$ couple to their conjugates, the relaxation rates for these density matrix elements are unchanged, but the oscillation frequencies are modified [see Eq.~(\ref{eigenvalue4})].  The relaxation rate for $\Delta\rho$ is coupled to $\ket{1}\bra{-1}$.  There are one real eigenvalue and a pair of complex conjugate eigenvalues which are different from non-strain case.  The former mode is localized to the coordinate $\Delta\rho$, and the later are localized to $\ket{1}\bra{-1}$ and $\ket{-1}\bra{1}$.

\section{Radio Frequency Field}  \label{Sec:rf}

\begin{figure}
\begin{center}
\includegraphics[width=0.55\textwidth]{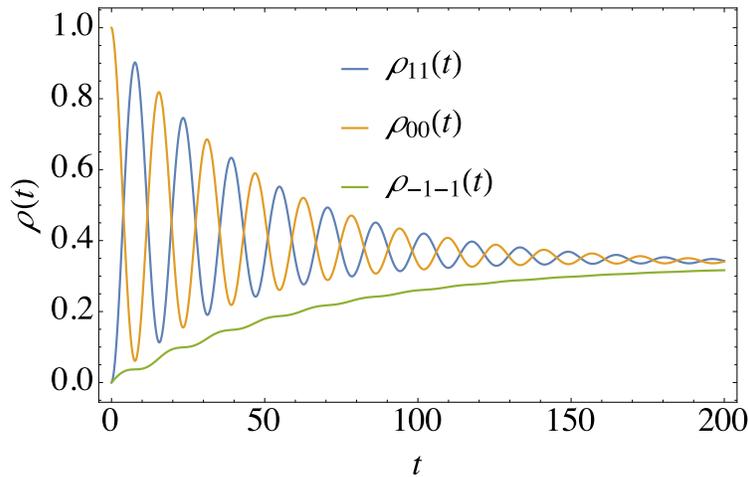}
\caption{(Color online) Population dynamics in the presence of a resonant rf field, $\rho_{11}(t)$ (blue), $\rho_{00}(t)$ (orange), $\rho_{-1 -1}(t)$ (green), calculated using the RWA.}
\label{Fig:resonance123}
\end{center}
\end{figure}

\begin{figure}
\begin{center}
\includegraphics[width=0.55\textwidth]{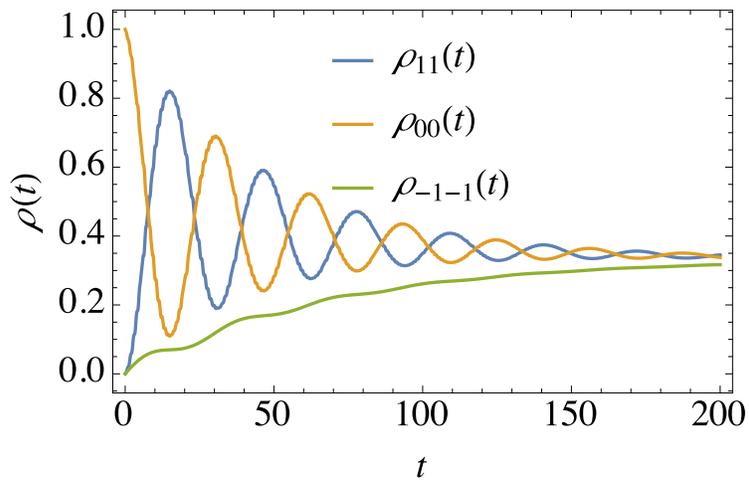}
\caption{(Color online) Population dynamics in the presence of a resonant rf field {\it without} making the RWA, $\rho_{11}(t)$ (blue), $\rho_{00}(t)$ (orange), $\rho_{-1 \, -1}(t)$ (green).}
\label{Fig:WithoutRWA}
\end{center}
\end{figure}

Additional information about NV centers can be gleaned from experiments in the presence of a radio frequency field tuned to frequencies close to resonance with a ground state transition, e.g., the $m = 0 \to m = -1$ transition.  We now consider the effects of a radio frequency field that is in or near resonance with an energy difference between two of the levels of the three-level system, e.g., $\omega_{\mathrm{rf}} \approx {\cal D} - \mu B_z$.  Experimentally, such transitions have been extensively studied~(see \cite{Lange12,Pla13} and references therein).  Figures~\ref{Fig:resonance123}  and \ref{Fig:WithoutRWA} show the population dynamics in the presence of a rf field starting from a ground state, i.e. $\rho_{\mathrm{initial}}=\ket{0}\bra{0}$, calculated with and without making the rotation wave approximation (RWA).  In this on resonance case, the RWA reproduces the dynamics of the non-RWA density matrix except for tiny high-frequency oscillations.  Therefore, below we study in greater detail the rf case within the RWA.  After making the rotating wave approximation, the time-independent Hamiltonian, including the rf coupling Hamiltonian $H_{\mathrm{rf}}$, is
\begin{eqnarray*}  \label{Eq:H_RWA}
H_{\mathrm{RWA}} &=& H + H_{\mathrm{rf}} = 
		\left(
		\begin{array}{ccc}
			\omega_1 & 0 & \epsilon^*\\
			0 & \omega_0 & 0\\
			\epsilon & 0 & \omega_{-1}		
		\end{array}
		\right) +
		\left(
		\begin{array}{ccc}
			0 & \Omega & 0 \\
			\Omega & \omega_{\mathrm{rf}} & 0 \\
			0 & 0 & 0 \\
		\end{array}
		\right) .
\end{eqnarray*}
Thus, the dressed state energy of the $m=0$ state is $\omega_{\mathrm{rf}}$ and the off-diagonal matrix element given by the Rabi frequency $\Omega$ couples $m=1$ and $m=0$.  Thus, contrary to the case without a rf field, the diagonal elements $\rho_{11}$ and  $\rho_{00}$ of the density matrix are now coupled to $\rho_{10}$ and it is no longer possible to separate Hilbert space as before.  Hence, even $\rho_{00}$ contains all 9 eigenmodes, in contrast with the no-rf case where $\rho_{00}$ contains only one eigenmode.

\begin{figure}
\begin{center}
\includegraphics[width=0.55\textwidth]{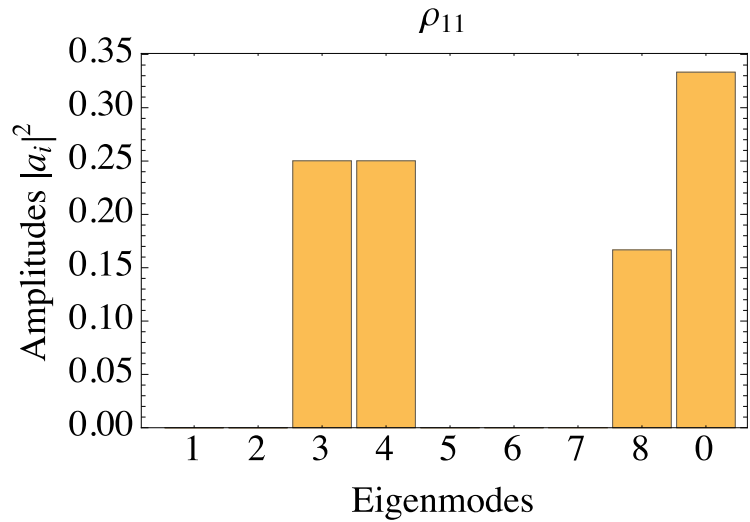}
\caption{(Color online) Distribution of the eigenmodes 0 through 8 in $\rho_{11}(t)$.}
\label{Fig:ini1}
\end{center}
\end{figure}

\begin{figure}
\begin{center}
\includegraphics[width=0.55\textwidth]{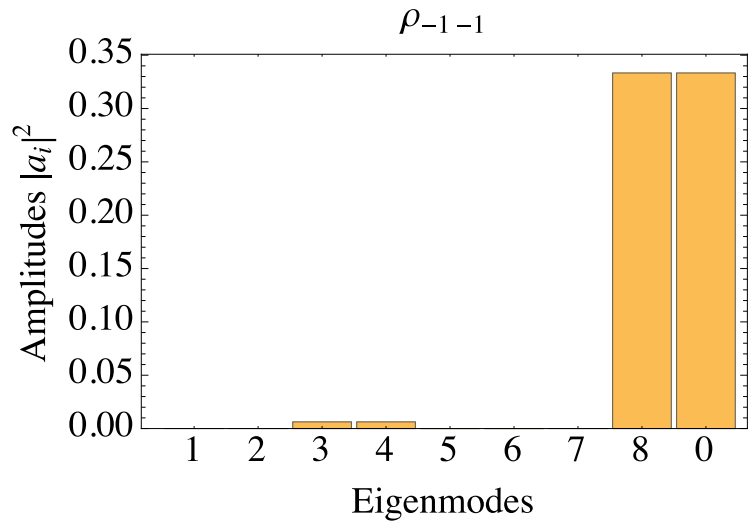}
\caption{(Color online) Distribution of the eigenmodes 0 through 8 for $\rho_{-1 \, -1}$.}
\label{Fig:ini3}
\end{center}
\end{figure}

Using a numerical diagonalization, we solved the ordinary differential equation with the following parameters: ${\cal D} = 2.87$ GHz, $\mu B_z=0.2$ GHz, $\Omega=0.2$ GHz, $\epsilon_x=\epsilon_y=0.0$, and $\omega_{\mathrm{rf}} = 2.87 + 0.2$ GHz.  When no rf field is present, the effect of the various time scales (i.e., the inverse of the decay rates) is separated into different subspaces, whereas, with an rf field present, the dynamics of each of the density matrix elements includes all 9 modes (1 steady state and 8 decay modes).  As discussed in Sec.~2, any of the elements of the density matrix can be expressed in the form of $\rho_{\alpha \beta}(t) = a^{ij}_0 +\sum_{i=1}^8 a^{\alpha \beta}_i \exp (\lambda_i t)$ [see Eq.~(\ref{Eq:rho_expand})].  However, it might be difficult to extract 9 modes from experimental data.  To determine the ideal functional form for fitting experimental data, we fit the numerically computed density matrix elements $\rho_{ij} (t)$ with three modes and compare the fit with the calculated results.  The distributions of modes for $\rho_{11}$ and $\rho_{-1 \, -1}$ are shown in Fig.~\ref{Fig:ini1} and Fig.~\ref{Fig:ini3}, where the vertical axis is $|a_i|^2$ and the horizontal axis is the mode index ($\rho_{00}$ is almost identical to $\rho_{11}$).  In  Fig.~\ref{Fig:ini1} and Fig.~\ref{Fig:ini3}, the first through the 6th modes correspond to three pairs of complex conjugate modes, the 7th and 8th modes are real, and the last mode (mode 0) corresponds to the steady state.

\begin{figure}
\begin{center}
\includegraphics[width=0.55\textwidth]{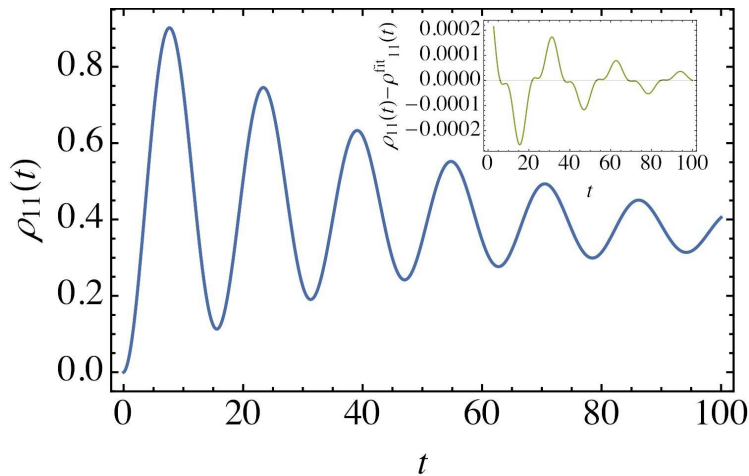}
\caption{(Color online) Comparison of $\rho_{11}(t)$ and the fit $\rho^{\mathrm{fit}}_{11}(t)$.  No visible difference between the solution and the fit can be seen.  The inset shows the difference $\rho_{11}(t)-\rho^{\mathrm{fit}}_{11}(t)$.  Note the scale of the ordinate of the inset is more than three orders of magnitude smaller than that of the main figure.}
\label{Fig:resonance1fit}
\end{center}
\end{figure}

The diagonal elements are dominated by the steady state, one relaxation mode without oscillation, and a pair of oscillating relaxation modes.
Thus, we fit $\rho_{11}$, $\rho_{00}$, and $\rho_{-1 \, -1}$ with the functional form
\begin{equation}
\rho_{\alpha \alpha}^{\mathrm{fit}}(t) = a^{\alpha} + b^{\alpha} \, e^{-\gamma_1^{\alpha} t} + \big[c^{\alpha} \, \cos(\omega_1^{\alpha} t) + d^{\alpha} \, \sin(\omega_1^{\alpha} t) \big] e^{-\gamma_2^{\alpha} t} .
\end{equation}
Fig.~\ref{Fig:resonance1fit} shows a comparison between $\rho_{11}$ and $\rho^{\mathrm{fit}}_{11}$ for the on-resonance rf calculations.  Although only three modes are considered in $\rho_{11}^{\mathrm{fit}}$, $\rho_{11}$ is reproduced almost perfectly by the fitting function $\rho^{\mathrm{fit}}_{11}(t)$.  We also found that the fits for $\rho_{00}$ and $\rho_{-1 \, -1}$ are good.  We note that, depending on the initial state at $t = 0$, the density matrix elements $\rho_{ij}(t)$ can contain different modes.  Sometimes it contains two real decay modes and two complex modes, and sometimes four complex modes, {\it etc}.  One must check to see how many modes are necessary in the fitting function and whether they are real or complex in order to fit experimental signals.

\begin{figure}
\begin{center}
\includegraphics[width=0.55\textwidth]{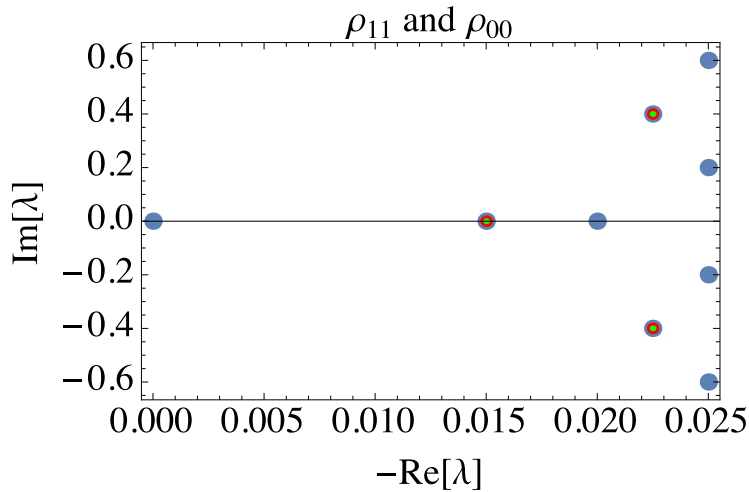}
\caption{(Color online) The 9 eigenmodes (blue points), and the 3 modes obtained by fitting [resonant case with $\omega_{\mathrm{rf}} = (2.87 + \mu \, B_z)$ GHz] for $\mu B_z = 0.2$ GHz. Red points are obtained from $\rho_{11}$ and green points are obtained from $\rho_{00}$.}
\label{Fig:eigen12}
\end{center}
\end{figure}

\begin{figure}
\begin{center}
\includegraphics[width=0.55\textwidth]{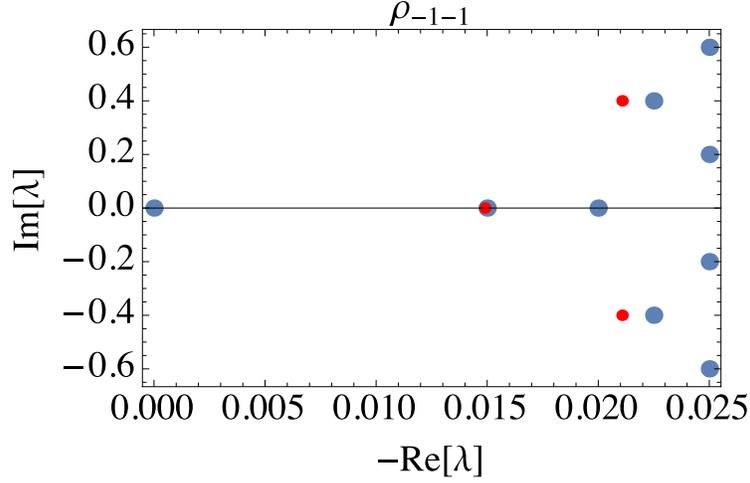}
\caption{(Color online) 9 eigenmodes, and the 3 modes obtained by fitting. Red points are obtained from $\rho_{-1 \, -1}$.}
\label{Fig:eigen3}
\end{center}
\end{figure}

\begin{figure}
\begin{center}
\includegraphics[width=0.55\textwidth]{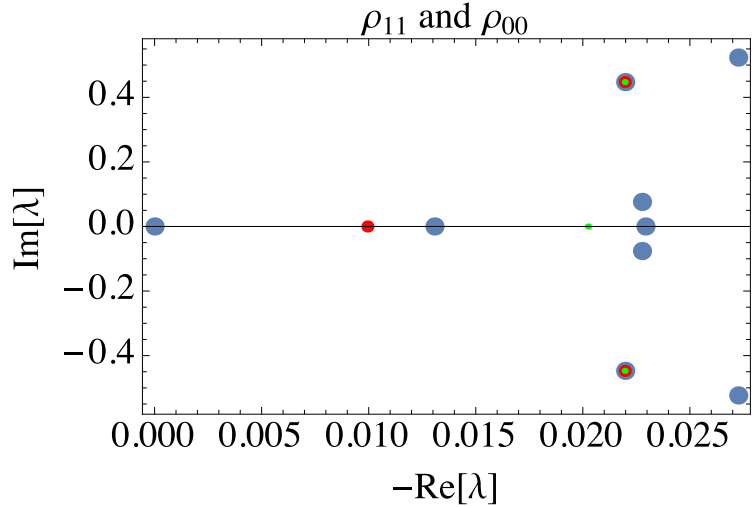}
\caption{(Color online) The 9 eigenmodes, and the 3 modes obtained by fitting (off-resonant case $\omega_{\mathrm{rf}}=2.87$ GHz) for $\mu B_z = 0.2$ GHz. Red points are obtained from $\rho_{11}$ and green points are obtained from $\rho_{00}$.}
\label{Fig:eigen12off}
\end{center}
\end{figure}

To determine the difference between the decay modes obtained in the calculated results and the three-mode fit, we plotted the 9 eigenmodes and the 3 mode fit of $\rho_{11}$ and $\rho_{00}$ in Fig.~\ref{Fig:eigen12}.  For the given parameters, the three extracted decay modes are {\it almost exactly the same} as the 9 eigenmodes calculated by diagonalization.  However, this is not always the case.  Sometimes there are small shifts of the decay modes compared to the exact eigenmodes. Therefore, we suspect that the decay time backed out of experimental data can be different from the true eigenvalues of the system.  For example, Fig.~\ref{Fig:eigen3} shows the 9 eigenmodes and the modes obtained by fitting $\rho_{-1 \, -1}$ with $\rho^{\mathrm{fit}}$ (the red curve in the figure). Even though $\rho_{-1-1}$ can be fit without visible error, the real eigenvalue has a small shift and a pair of complex eigenvalues has a somewhat larger shift.  Note that this is not a special property for an uncoupled state~$\rho_{-1-1}$.  Such a shift may occur for any of the matrix elements.  For example, let us take an radio frequency that is 10\% smaller than the resonance frequency ($\omega_{\mathrm{rf}}=2.87$ GHz).  In this case, the time evolution was perfectly reproduced as in the case of resonant rf frequency ($\omega_{\mathrm{rf}} = 3.07$ GHz) but we find a small shift of decay modes for $\rho_{11}$ and $\rho_{00}$ (see Fig.~\ref{Fig:eigen12off}).

To summarize the discussion in this section, in presence of a rf field that couples two of the levels of the three-level system, many modes contribute to each density matrix element.  We have examined the distribution of eigenmodes obtained by diagonalization and fit the calculated density matrix elements with a simple fitting function.  Our analysis could be easily generalized to study more complicated systems involving more spins (e.g., two nearby NVs, a nitrogen impurity near an NV, or a ${}^{13}$C impurity near an NV \cite{Childress06}) in order to extract essential frequencies of time evolution in a systematic way.

\section{Coherence Relaxation} \label{Sec:Decoherence}

Decay of the coherences of NVs, i.e., of the off-diagonal elements of the density matrix, also occurs due to interaction with the environment.  The decay time for the coherences is often referred to as $T_2$, i.e., the ``transverse spin relaxation time'', but for the three-level NV system, there can be multiple coherence decay time scales.  In this section we present results for the decay of the $\rho_{1 0}(t)$ density matrix element in the presence of a resonant rf field.  Our results are analytic, except for the diagonalization of a 9$\times$9 matrix.
Then, we fit the analytically determined decay curves to simple forms with the hope of learning what a fit to experimental data might yield.  Coherence decays are often experimentally studied by following the Rabi oscillations of electromagnetically coupled levels.  Dynamical decoupling pulse sequences can extend the transverse spin relaxation time for systems that are inhomogeneously broadened, thereby improving sensitivity and coherence time \cite{t2}.

\begin{figure}
\centering\subfigure[]{\includegraphics[width=0.45\textwidth]
{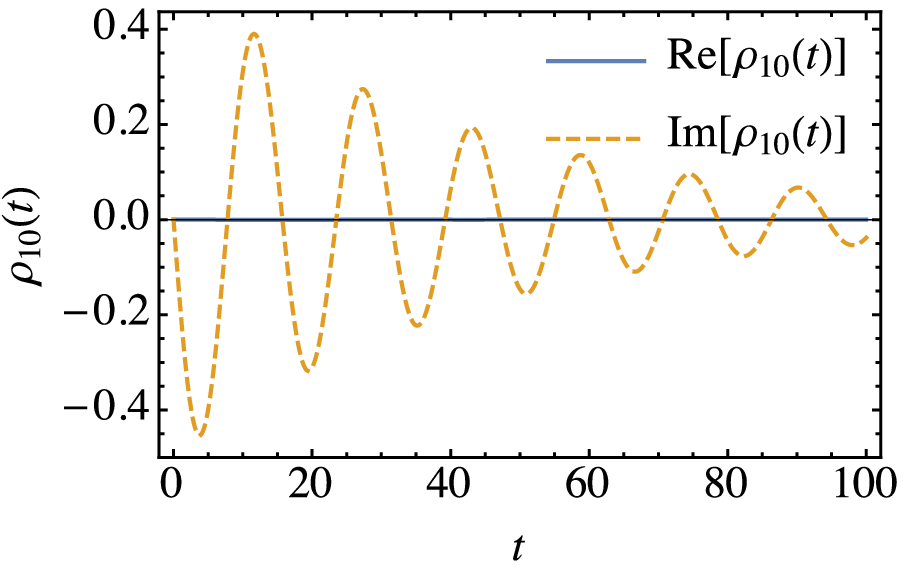}}
\centering\subfigure[]{\includegraphics[width=0.45\textwidth]
{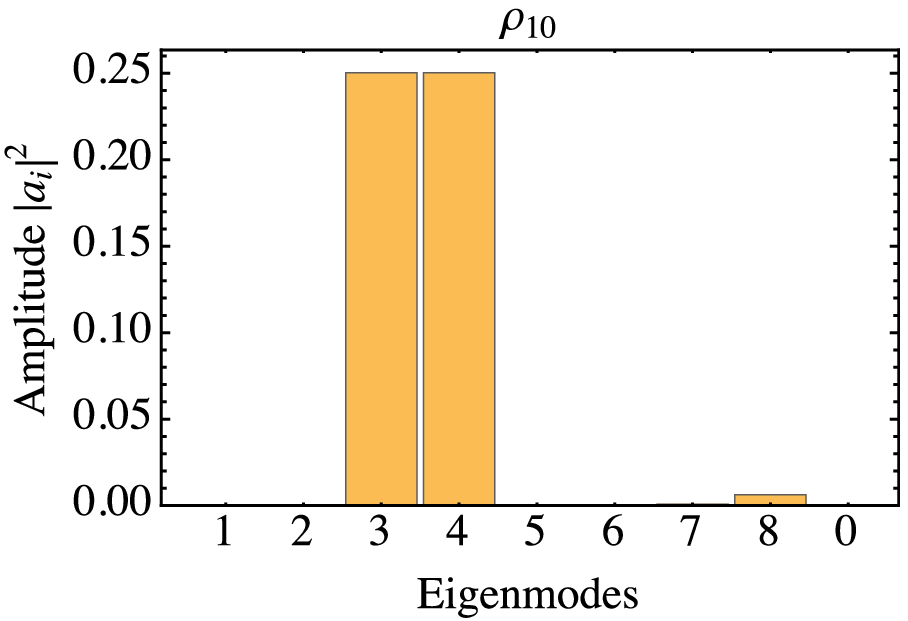}}
\centering\subfigure[]{\includegraphics[width=0.45\textwidth]
{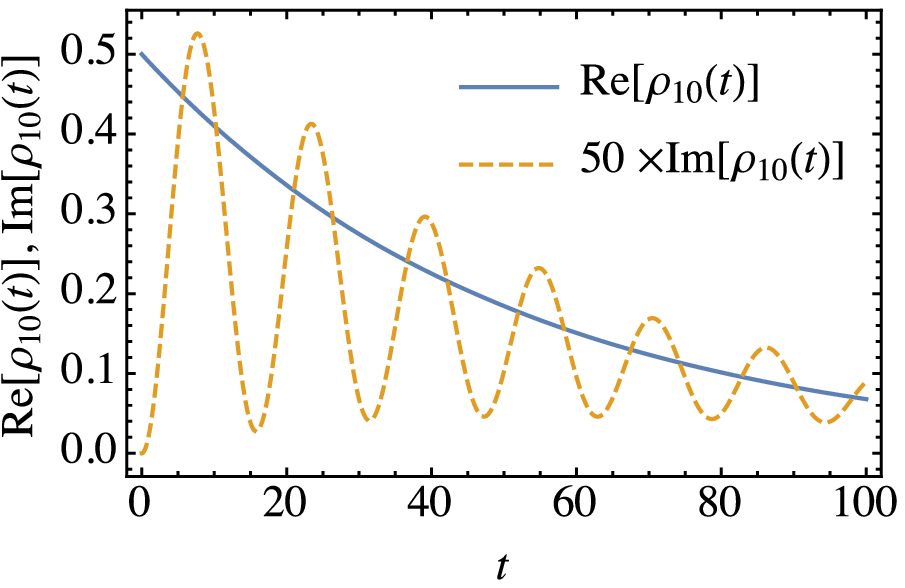}}
\centering\subfigure[]{\includegraphics[width=0.45\textwidth]
{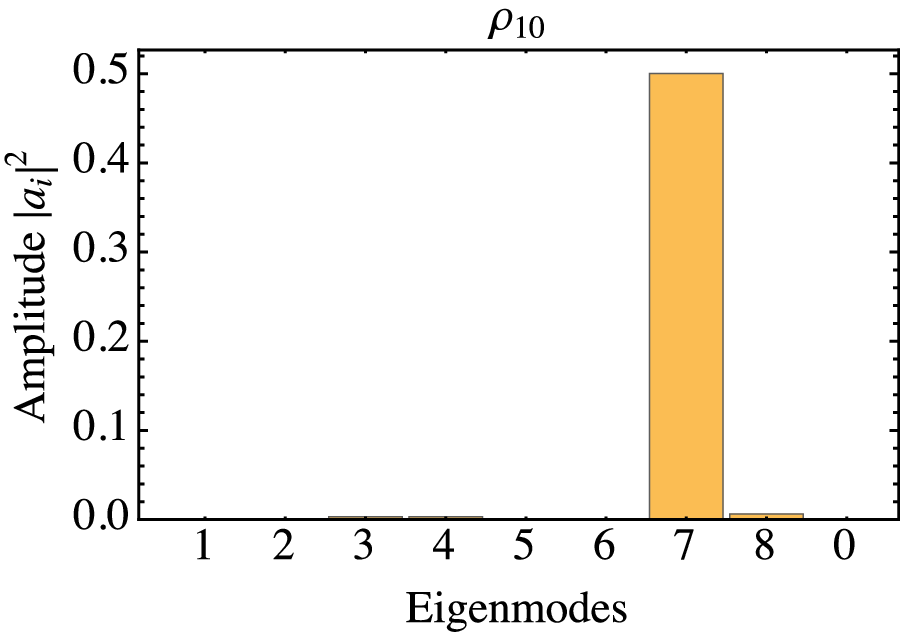}}
\caption{(Color online) Coherence relaxation in the presence of a rf field.  (a) The dynamics of the off-diagonal density matrix element $\rho_{1 0}(t)$ in the presence of a resonant rf field when the initial state is taken as the ground state.  (b) Distribution of the eigenmodes 0 through 8 for $\rho_{1 0}$ shown in (a).  (c) The dynamics of the off-diagonal density matrix element $\rho_{1 0}(t)$ in the presence of a resonant rf field when the initial state is taken as $\psi_{\mathrm{in}}(0) = \frac{1}{\sqrt{2}} (|m=0\rangle + |m=1\rangle)$.  (d) Distribution of the eigenmodes 0 through 8 for $\rho_{1 0}$ shown in (c).}
\label{Fig:decoherence}
\end{figure}

We calculate the dynamics of an NV center in the presence of a resonant rf field when the initial state is taken as the ground state, $\rho_{\rm{in}} = \rho_{ij}(0) = \delta_{i0} \delta_{j0}$, and when the initial state is taken a a coherent superposition of $m = 0$ and $m = 1$, $\psi_{\mathrm{in}}(0) = \frac{1}{\sqrt{2}} (|m=0\rangle + |m=1\rangle)$.  Figure~\ref{Fig:decoherence} plots the time dependence of the off-diagonal density matrix element $\rho_{1 0}(t)$ and the distribution of the eigenmodes in the density matrix element for both initial conditions.  The time dependence of the real and imaginary parts of $\rho_{1 0}(t)$ for the initial state being the ground state is shown in Fig.~\ref{Fig:decoherence}(a).  The real part of $\rho_{1 0}(t)$ is multiplied by a factor of 200 to make it visible in the figure.  Fig.~\ref{Fig:decoherence}(b) shows that the 3rd, 4th and 8th eigenmodes substantially contribute to $\rho_{1 0}(t)$.  The square of the amplitudes for the modes are: $|a_i|^2 = 0.0001168, 0.0000334, 0.250352, 0.250268, 0.000279, 0.0001391, 0.0008345, 0.0062619$ for $i = 1, 2, 3, 4, 5, 6, 7, 8$.  Figure~\ref{Fig:decoherence}(c) plots the time dependence of the off-diagonal density matrix element $\rho_{1 0}(t)$ when the initial state is $\psi_{\mathrm{in}}(0) = \frac{1}{\sqrt{2}} (|0\rangle + |1\rangle)$, and Fig.~\ref{Fig:decoherence}(d) shows that eigenmode 7 is the largest populated mode, eigenmodes 3, 4 and 8 are present with smaller amplitudes, and the other modes are even smaller.  The square of the amplitudes for this case of initializing to a coherent state are: 0.0000973, 0.00002782, 0.003123, 0.00312244, 0.0004170, 0.0002080, 0.500276, 0.0062805, respectively.

When the initial state is taken as the ground state, a numerical fit of the coherences $\rho_{1 0}(t)$, $\rho_{1 \, -1}(t)$ and $\rho_{0 \, -1}(t)$, proved to be difficult, but when the initial state is taken to be the coherent state, we could fit well with the fitting function,
\begin{equation} \label{fit_rho_10}
{\mathrm{Re}}[\rho_{10}^{\mathrm{fit}}(t)] = b_r \, e^{-\gamma_r t} + \big[c_r \, \cos(\omega_r t) + d_r \, \sin(\omega_r t) \big] e^{-\Gamma_r t} , \quad {\mathrm{Im}}[\rho_{10}^{\mathrm{fit}}(t)] = b_i \, e^{-\gamma_i t} + \big[c_i \, \cos(\omega_i t) + d_i \, \sin(\omega_i t) \big] e^{-\Gamma_i t} .
\end{equation}
The optimal values of the coeficients in Eq.~(\ref{fit_rho_10}) are:
$b_r =  0.500676$, $c_r = -0.000627388$, $d_r =  -2.57696 \times 10^{-6}$, $\gamma_r = 0.0199914$, $\omega_r = 0.199101$, and $\Gamma_r= 0.0251007$, and $b_i =  0.00619681$, $c_i = -0.00612731$, $d_i =   -0.000198632$, $\gamma_i = 0.0148699$, $\omega_i = 0.399882$, and $\Gamma_i= 0.0221052$, respectively.  The $\Gamma_{r,i}, \gamma_{r,i}$ and $\omega_{r,i}$ correspond to 3rd through the 8th eigenmodes.  The coherences $\rho_{1 \, -1}(t)$ and $\rho_{0 \, -1}(t)$ when the initial state is taken to be the coherent state can be used to numerically determine eigenmodes 1 and 2.

\section{Summary and Conclusions}   \label{Sec:summary}

We discussed the dynamics of NV centers in diamond in the presence of a fluctuating magnetic field that arises from the surroundings (the environment) with and without the presence of a rf field that couples levels of the ground electronic state manifold. The NV ground state is a triplet, hence three levels play a role in the dynamics and we studied its decoherence and dephasing.  When neither strain nor a rf field are included in the description, the relaxation rate of the diagonal elements of density matrix towards a Boltzmann distribution and the relaxation of the off-diagonal elements (loss of quantum coherence) are easy to obtain analytically.  With strain and without rf field, $\rho_{00}$ does not couple to the other element and the dynamics is the same as the non-strain case.  The dominant modes for $\rho_{11}$, $\rho_{-1 \, -1}$, $\rho_{1 \, -1}$ and $\rho_{-1 \, 1}$ are affected by strain.  Surprisingly, we were able to analytically prove that even though $\rho_{10}$, $\rho_{0 \, -1}$, $\rho_{0 \, 1}$, and $\rho_{-1 \, 0}$ couple each other, the dominant modes for them have exactly the same relaxation rate as the non-strain case; strain only modifies the frequencies of their oscillation (see Sec.~\ref{Sec:w_Strain}).

When a rf field between $m=0$ and $m=1$ is present, all the modes appear in the dynamics of each of the density matrix elements, therefore, the conventional two-level system analysis for NVs can be very misleading.  Depending on the parameters used, we observe that the relaxation times obtained by the fit can be slightly different from those of the eigenmodes.  Hence, in order to understand the relaxation processes induced by a complicated environment and obtain accurate determination of relaxation times, one should use all the eigenmodes.

Before closing, let us comment on the decay dynamics at finite temperature. There are several ways that temperature can affect the NV system.  The strain parameter $\epsilon$ depends on the crystal phonons (and the electric field at the location of the NV), hence the phonon temperature affects of strain.  Also, the vibrational dynamics of the spins that produce the fluctuating magnetic field at the location of the NV affect the dipole-dipole interactions of the NV dipole moment and the other dipole moments in the crystal, hence the fluctuating field is affected by the phonons because they modify the relative coordinate vectors from the NV to the other dipole moments (these coordinate vectors appear in the dipole-dipole interaction).  The equilibrium state of the NV should also be temperature dependent (it should be a Boltzmann equilibrium state that depends on the temperature).  But, except at extremely low temperature, the Boltzmann factor in diamond NVs, $\approx\exp(-{\cal D}/k_B T)$, where $k_B$ is the Boltzmann constant, is almost unity; therefore, any temperature dependence of the equilibrium would be practically impossible to observe.  But for extremely low temperature, a departure of the equilibrium density matrix from the democratic state (the state with equal population in all levels) might be observable.  Temperature dependence is not explicitly built into our model.  One approach to include the effects of temperature is to derive a reduced density matrix equation from a microscopic model. To do so for NV centers, requires incorporation of the effects of phonons and spin baths.  After tracing out the spin bath and phonon degrees of freedom, one would obtain a master equation for the reduced density matrix.  However, this approach requires detailed knowledge for the interactions and the spectral density function of the baths~\cite{BreuerBook}.  Alternatively, one can consider less precise but phenomenological approaches to treat temperature effects.  One of the simplest approaches would be to add transition operators as Lindblad operators and artificially force the system to equilibrium at a given temperature by using asymmetric Lindblad coefficients.  For example, consider 6 transition operators between any two of the three levels of the NV ground state system.  If $\epsilon=0$, one can choose an arbitrary value for 4 of the 6 Lindblad coefficients (note that the 4 coefficients could be zero), and the choice of the 4 coefficients determines the other two coefficients (the other two Lindblad coefficients cannot be the pair $\Gamma_{ij}$ and $\Gamma_{ji}$).  For example, for the symmetric fluctuation case, $\gamma = \Gamma$, the other two Lindblad coefficients are obtained in the following manner.  Suppose we have Lidnblad coefficients $\Gamma_{01}$, $\Gamma_{0 \, -1}$, $\Gamma_{1 \, -1}$ and $\Gamma_{-11}$ from experiment, where $\sqrt{\Gamma_{ij}}$ is the Lindblad coefficient for the Lindblad operator $c^\dag_j c_i$.  Then, the remaining two coefficients are given by
\begin{equation}
\Gamma_{10} =
\frac{-(p_1-p_0) \Gamma + p_0 \Gamma_{01} + p_{-1} \Gamma_{-11} - p_1 \Gamma_{1-1}}{p_1} , \quad
\Gamma_{-10}  =
\frac{(p_0- p_{-1}) \Gamma - p_{-1} \Gamma_{-11} + p_1 \Gamma_{1-1} +  p_0 \Gamma_{0-1}}{p_{-1}}.
\end{equation}
where $p_m$ denotes a population of Boltzmann distribution at a given temperature for $(m = 0,\pm 1)$.  Although one still would need relaxation rates from experiment in order to determine the Lindblad coefficients, this approach is much simpler than determining the interaction and the spectral density functions of the baths.  Furthermore, if $\epsilon \neq 0$, one can choose arbitrary values for 3 of the 6 Lindblad coefficients, and the other 3 are completely specified.  In general, by adding coupling operators, one has less freedom to choose Lindblad coefficients.

\bigskip

{\it Acknowledgement.} This work was supported in part by grants from the Israel Science Foundation (Grant No.~295/2011) and the DFG through the DIP program (FO 703/2-1).  We are grateful to Professor Dmitry Budker for valuable discussions.

\bigskip

\section*{References}
\bibliographystyle{science}
{\footnotesize
\bibliography{NV_Spin1_System}}

\end{document}